\documentclass[12pt,letterpaper]{article}

\usepackage{natbib,hyperref}

\usepackage[ left=1in, top=1in, right=1in, bottom=1in]{geometry}
\usepackage{graphicx,bm,colonequals,amsmath,amssymb,url,xcolor,bbm}
\usepackage{array,tabularx,multirow}
\usepackage[font={footnotesize}]{caption,subcaption}
\usepackage[utf8]{inputenc}
\usepackage{enumitem}
\usepackage{soul} 
\usepackage{placeins} 
\usepackage[normalem]{ulem}

\usepackage{xr}
\makeatletter
\newcommand*{\addFileDependency}[1]{
  \typeout{(#1)}
  \@addtofilelist{#1}
  \IfFileExists{#1}{}{\typeout{No file #1.}}
}
\makeatother
\newcommand*{\myexternaldocument}[1]{%
    \externaldocument{#1}%
    \addFileDependency{#1.tex}%
    \addFileDependency{#1.aux}%
}
\myexternaldocument{supplement}

\usepackage{mathtools}
\mathtoolsset{showonlyrefs} 

\everydisplay{\textstyle}

\bibpunct[, ]{(}{)}{;}{a}{,}{,}
   
\usepackage{amsthm}
\newtheoremstyle{propstyle} 
    {2mm}                    
    {1mm}                    
    {\itshape}                   
    {}                           
    {\scshape}                   
    {.}                          
    {.5em}                       
    {}  
\theoremstyle{propstyle}

\theoremstyle{propstyle}

\theoremstyle{propstyle}

\theoremstyle{propstyle}

\theoremstyle{propstyle}

\makeatletter
\renewcommand{\paragraph}{%
  \@startsection{paragraph}{4}%
  {\z@}{2ex \@plus 1ex \@minus .2ex}{-1em}%
  {\normalfont\normalsize\bfseries}%
}
\makeatother

\DeclareMathAlphabet\mathbfcal{OMS}{cmsy}{b}{n}
\newcommand{\iid}{\stackrel{iid}{\sim}}
\newcommand{\indep}{\stackrel{indep}{\sim}}

\newcommand{\ba}{\mathbf{a}}

\newcommand{\bb}{\mathbf{b}}
\newcommand{\bs}{\mathbf{s}}

\newcommand{\bw}{\mathbf{w}}

\newcommand{\bX}{\mathbf{X}}

\newcommand{\bC}{\mathbf{C}}

\newcommand{\bfzero}{\mathbf{0}}

\newcommand{\bftheta}{\bm{\theta}}

\newcommand{\bfbeta}{\bm{\beta}}

\newcommand{\bfrho}{\bm{\rho}}
\newcommand{\bfsigma}{\bm{\sigma}}

\newcommand{\GP}{\mathcal{GP}}

\newcommand{\normal}{\mathcal{N}}

\newcommand{\domain}{\mathcal{D}}

\title{A nonstationary spatial model of PM$_{2.5}$ with localized transfer learning from numerical model output}
\date{}
\author{Wenlong Gong  \thanks{Department of Mathematics and Statistics, University of Houston - Downtown} 
\thanks{Corresponding author: \texttt{wenlonggong@gmail.com}}
\and Brian J. Reich \thanks{Department of Statistics, North Carolina State University}  
\and Joseph Guinness \thanks{Department of Statistics and Data Science, Washington University in St. Louis}}

\begin{document}

\maketitle

\begin{abstract}
Ambient air pollution measurements from regulatory monitoring networks are routinely used to support epidemiologic studies and environmental policy decision making. However, regulatory monitors are spatially sparse and preferentially located in areas with large populations. Numerical air pollution model output can be leveraged into the inference and prediction of air pollution data combining with measurements from monitors. Nonstationary covariance functions allow the model to adapt to spatial surfaces whose variability changes with location like air pollution data. In the paper, we employ localized covariance parameters learned from the numerical output model to knit together into a global nonstationary covariance, to incorporate in a fully Bayesian model. We model the nonstationary structure in a computationally efficient way to make the Bayesian model scalable.	\\


\end{abstract}



\section{Introduction \label{sec:intro}}

Ambient air pollution poses a significant threat to public health and contributes to substantial global health costs. Among pollutants, fine particulate matter with a diameter of 2.5 microns or less (PM$_{2.5}$) is particularly hazardous, being linked to a wide range of adverse health effects, from minor respiratory irritation to elevated premature mortality \citep{Bernstein2004, Kampa2008,Kaufman2016,Pope2020}. Accurate and reliable exposure estimates are essential for successful environmental health studies and policy development.\\ 

Regulatory monitoring networks provide direct measurements of air pollution and plays a critical role in supporting epidemiological research and informing environmental policies. However, their sparse spatial coverage, particularly outside densely populated urban areas, creates large gaps in data, which complicating efforts to assess the true spatial distribution of PM$_{2.5}$ and its impacts on diverse communities.\\

Numerical air quality models, which simulate air quality based on the chemical and physical dynamics of pollutants, offer a comprehensive large-scale view of air pollution concentrations. For example, the Community Multiscale Air Quality Modeling (CMAQ) system developed by the U.S. Environmental Protection Agency (EPA) provides regional-scale estimates of air quality. While such models offer comprehensive spatial coverage, they require calibration and validation against monitoring data to ensure accuracy. Integrating numerical model outputs with observations has become a central strategy in environmental statistics, with approaches including data assimilation, which sequentially updates model states with new information \citep{Wikle2007, Wikle2019}; Bayesian melding and hierarchical modeling, which provide coherent uncertainty quantification across data sources \citep{Poole2000, Reich2011, Chang2015}; statistical data fusion \citep{Fuentes2005, Berrocal2010}, which merges multiple monitoring and model products ; downscaling \citep{McMillan2010, Berrocal2012, pierce2023, Zheng2025}, which translates coarse-resolution outputs to local scales ; and machine and deep learning methods, which enable flexible bias correction and scalable modeling of high-dimensional fields \citep{Katzfuss2024, Fang2025}. Complementary efforts have also focused on calibrating and validating numerical models using supplemental monitoring networks, particularly through regression, ensemble, and spatial filtering frameworks \citep{Chu2016, Yu2018, Chu2020, Casciaro2022, Heffernan2023, Chen2024}.. \\ 

Building on this body of work, our study focuses on modeling PM$_{2.5}$ concentration from monitoring data while leveraging numerical model outputs to improve spatial prediction coverage and uncertainty quantification. Specifically, we use empirical orthogonal functions to capture large-scale patterns from CMAQ, and a spatial model for local variations from monitoring data. Empirical orthogonal function (EOF) analysis effectively characterizes dominant patterns in complex space–time processes by decomposing the covariance kernel into eigenfunctions, reducing the dimensionality of space–time random fields into a smaller set of new spatial fields \citep{Storch1999, Hannachi2007}. The mean effects are represented as a linear combination of the derived EOFs, capturing spatial patterns at various scales. In the spatial model of local variation, we incorporate a transfer learning perspective \citep{Pan2010}, enabling information from the numerical model domain to inform parameter estimation in the observational domain.\\

Transfer learning has become increasingly valuable in spatial modeling, particularly for enhancing prediction accuracy in regions with limited data. It has been applied in spatial context to improve model performance by transferring information from data-rich regions to data-sparse areas. For example, it has been used in spatial autoregressive models and urban computing scenarios to borrow spatial dependencies from well-observed locations \citep{Wang2019, Shao2021, Zeng2024}. In our work, we adopt this concept by using CMAQ outputs as a reference field to estimate local spatial parameters, allowing the model to adapt to regional variability while maintaining computational efficiency. More critically, we model the spatial random effect as a nonstationary process, which is particularly suited for spatial surfaces with location-dependent variability — a common feature of air pollution data.\\

Existing methods for modeling nonstationary spatial processes include basis function expansions, process convolution, spatial deformation. Basis function expansions define nonstationary processes as Karhunen-Loeve expansion, while convolution models define them through convolution kernels. Other approaches include using wavelet bases \citep{Nychka2002}, empirical orthogonal functions basis set \citep{Wikle1999}, and random point process basis functions \citep{Katzfuss2012}. Although these methods offer computational efficiency, they often produce covariance functions as sums over numerous terms, which can complicate interpretation and hinder the representation of local spatial features. An alternative method for generating explicit covariance functions with locally varying geometric anisotropies is spatial deformation \citep{Sampson1992,Schmidt2003, Perrin2000,Clerc2003}.  \citet{Higdon1998} created a nonstationary version covariance function through the convolution of spatially-varying kernels with a stationary covariance function. Similarly, \citet{Fuentes2002} produced nonstationary covariance functions by convolving a fixed kernel over independent stationary processes with varying covariance parameters. These convolution methods often require challenging numerical calculations of integrals. \citet{Paciorek2006} addressed this by introducing explicit expressions for a class of nonstationary covariance function with locally varying parameters, derived from analytic solutions of the convolution integral. \citet{Stein2005} further demonstrated that, in addition to locally varying anisotropies, spatial smoothness can also vary in space based on this framework. Further extensions have been developed to allow spatial or spatio-temporal covariance functions to vary with covariates. For example, \citet{Calder2008} and \citet{Risser2016} incorporated covariates to the convolution-based covariance kernel, while \citet{Schmidt2011} used covariates as augmented coordinates in the deformation approach. \citet{Reich2011} proposed a model that combines stationary processes with coefficients that vary based on space-time covariates. More recently, \cite{chin2024} integrates covariates like wind speed into covariance function through a Cholesky-type decomposition, applied to PM$_{2.5}$ data.\\


In our approach, we specify a non-stationary Mat\'ern covariance for the spatial random effect, use CMAQ numerical model outputs as reference fields to derive local parameters. A key advantage of using CMAQ data is its comprehensive spatiotemporal coverage, which facilitates the generation of EOFs and allows the estimation of local parameters across multiple time points. Additionally, CMAQ data captures broader spatial patterns, especially in rural areas not covered by monitoring stations. By leveraging localized covariance parameters from numerical model outputs, we construct a global nonstationary covariance framework. This framework is incorporated into a fully Bayesian model to enable uncertainty quantification, with the latent non-stationary process leveraging the estimated local spatial parameters. To reduce computational load, we use a simple linear fit for these parameters, based on estimates from numerical outputs, rather than estimating them through MCMC. This approach not only captures the nonstationary structure of air pollution but also enhances computational efficiency by reducing the number of parameters, making the Bayesian model scalable for large datasets. \\


The remainder of this article is organized as follows.  Section \ref{sec:data} describes of the air pollution data. Section \ref{sec:methodology} details the components of our model, including EOFs and our extended nonstationary spatial model. Sections \ref{sec:sim} and  \ref{sec:app} evaluate the model performance using simulated data and a real-world PM$_{2.5}$ data. Finally, conclusions and discussions are given in Section \ref{sec:conclusions}.\\




\section{Data \label{sec:data}}

We utilize two primary sources of ambient air pollution data: numerical model outputs and monitoring network observations.\\

\begin{figure}
\includegraphics[trim = 0cm 0cm 6.5cm 0cm, clip,width=1\textwidth]{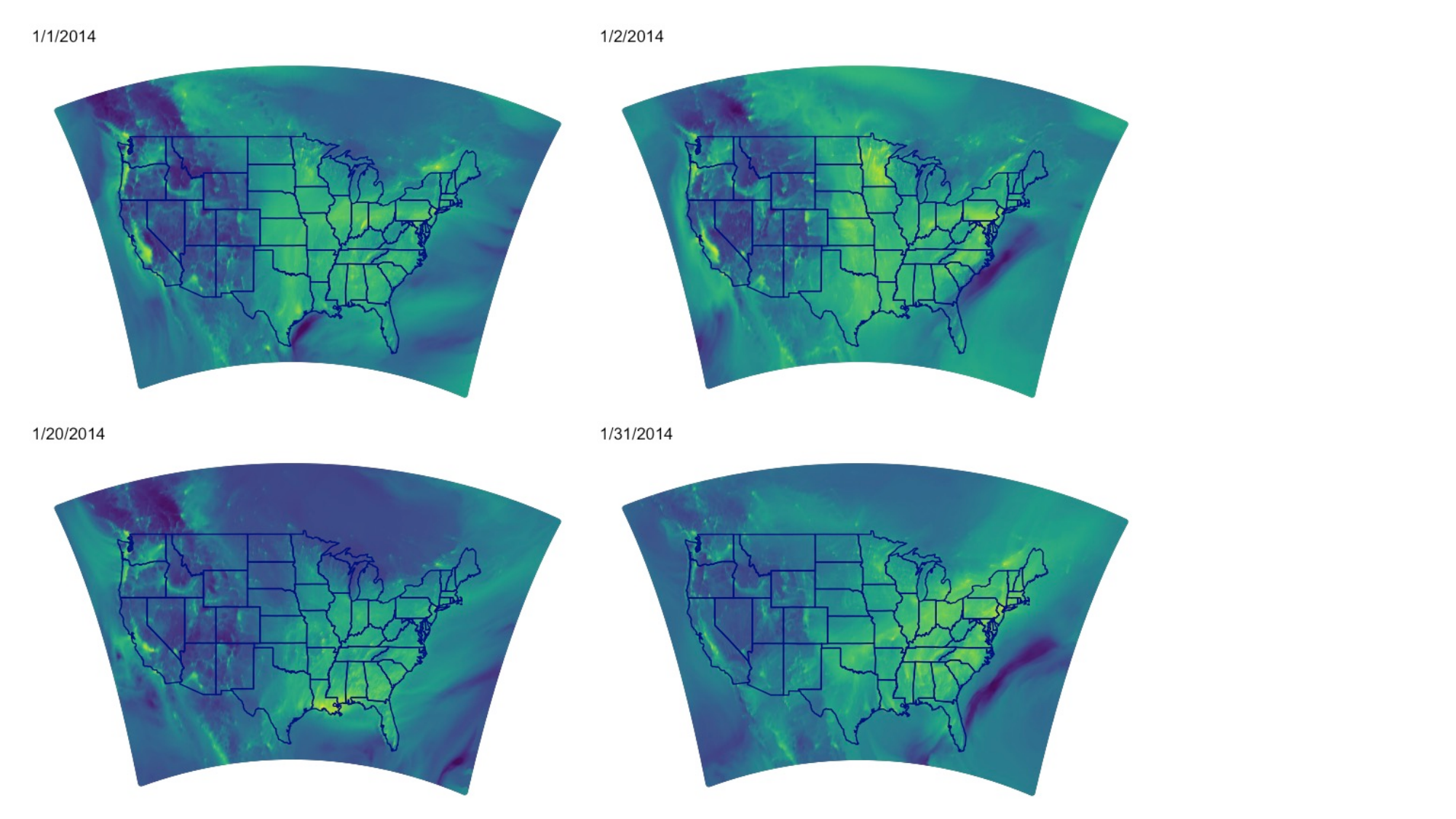} 
\caption{\label{fig:cmaq} CMAQ output examples of PM$_{2.5}$ in log scale of $\mu g/ m^3$.}
\end{figure}

The first source is the output from the Community Multiscale Air Quality (CMAQ) model, a 3-dimensional chemical transport model (CTM) developed by the U.S. Environmental Protection Agency (EPA). CMAQ integrates meteorological data, emission inventories, and atmospheric chemical and physical processes to simulate pollution concentrations at a large scale \citep{Binkowski2003, Appel2008}. The key advantage of using CTM outputs, such as those from CMAQ, is their complete spatiotemporal coverage, which allows for continuous air quality predictions across broad regions and time frames. Figure \ref{fig:cmaq} illustrates a four-day snapshot of PM$_{2.5}$ concentrations simulated by CMAQ. And the plot was made with R package \texttt{ggplot2} \citep{ggplot2}. \\


\begin{figure}
\centering\includegraphics[trim = 0cm 0cm 0cm 0cm, clip, width=0.9\textwidth]{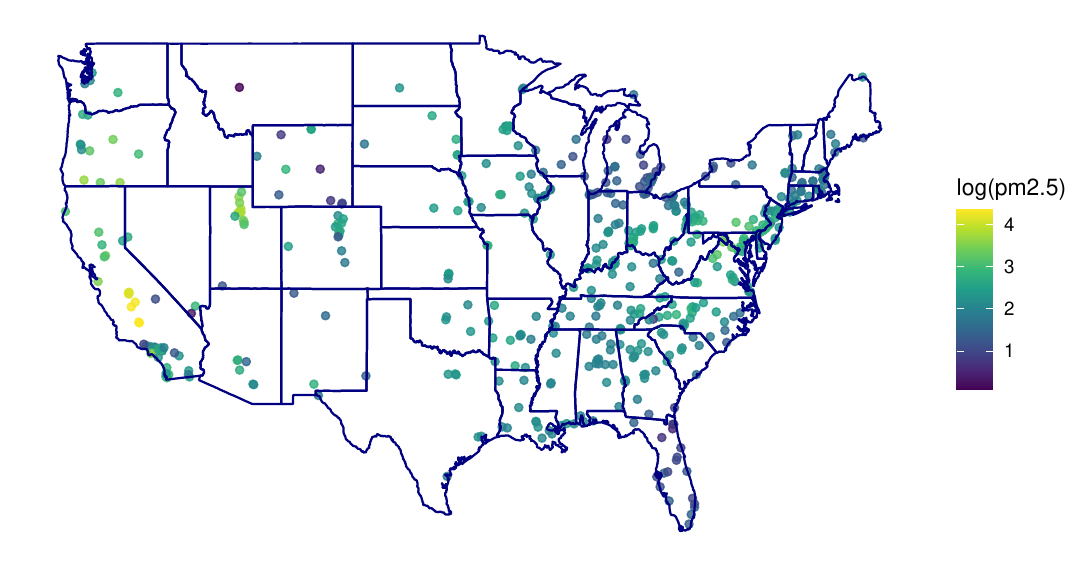} 
\caption{\label{fig:aqs} AQS observations of PM$_{2.5}$ (log scale of $\mu g/ m^3$) on January 2, 2014 .}
\end{figure}

The second source is observational data from the Air Quality System (AQS), which compiles air pollution measurements from a network of monitors operated by the EPA, state, local, and tribal air agencies. This dataset provides real-world ambient air pollution data collected at specific monitoring sites. While these measurements are highly accurate, the spatial coverage of the monitors is limited, primarily focusing on densely populated urban areas. Figure \ref{fig:aqs} shows PM$_{2.5}$ concentration captured by monitors on a single day.\\

\section{Methodology \label{sec:methodology}}

The CMAQ data shown in Figure \ref{fig:cmaq} provides complete spatiotemporal coverage and displays the large-scale spatial patterns of the field. To capture these dominant patterns efficiently, we reduce the dimension of the space–time random fields into a smaller set of spatial fields that serve as explanatory variables in our model's main effect. Empirical Orthogonal Function (EOF) analysis is particularly effective for identifying these patterns in complex space–time processes.

\subsection{EOF}
Empirical orthogonal function (EOF) analysis is a powerful technique for identifying dominant patterns in complex space–time data, with early applications in atmospheric science \citep{Obukhov1960,  Kutzbach1967, Lorenz1956, Lorenz1970}. It decomposes a continuous space-time random field into a set of orthogonal spatial patterns (the EOFs) and their corresponding uncorrelated time series, known as principal components (PCs). EOF analysis is mathematically equivalent to principal component analysis (PCA) \citep{Monahan2009}, with the term “EOF” more commonly used in atmospheric and geophysical sciences.\\

Consider a space-time field $X$ observed at grid locations $s$ and discrete time $t$ with the observations denoted by $x_{t,s}$ where $t=1,\ldots, p$ and $s=1,\ldots, n$. The observed field can be written as a data matrix
\[
\bX = \begin{pmatrix}
x_{1,1} & x_{2,1} & \ldots &x_{p,1}\\
x_{1,2} & x_{2,2} & \ldots &x_{p,2}\\
\ldots & \ldots & \ldots &\ldots\\
x_{1,n} & x_{2,n}  & \ldots &x_{p,n}
\end{pmatrix}.
\]

To center the data, we subtract the row-wise mean from each row, yielding the anomaly matrix \(\Tilde{\bX} \in \mathbb{R}^{n \times p}\). We then perform singular value decomposition (SVD), \(\Tilde{\bX} = E\Lambda U^T\), where the columns of \(E\) represent the empirical orthogonal functions (EOFs), orthogonal spatial patterns; and the columns of \(U\) are the corresponding principal components (PCs), representing temporal evolution. Because \(\Tilde{\bX}\) is structured with space in rows and time in columns, the EOFs capture dominant spatial variability, while the PCs reflect how these spatial patterns change over time. This decomposition provides an efficient and interpretable representation of the data, enabling dimension reduction by retaining only the first $M$ terms (where $M \ll r$, the rank of $\Tilde{\bX}$) \citep{Hannachi2001}.\\

\begin{figure}
\centering\includegraphics[trim = 0cm 0cm 0cm 1cm, clip, width=0.9\textwidth]{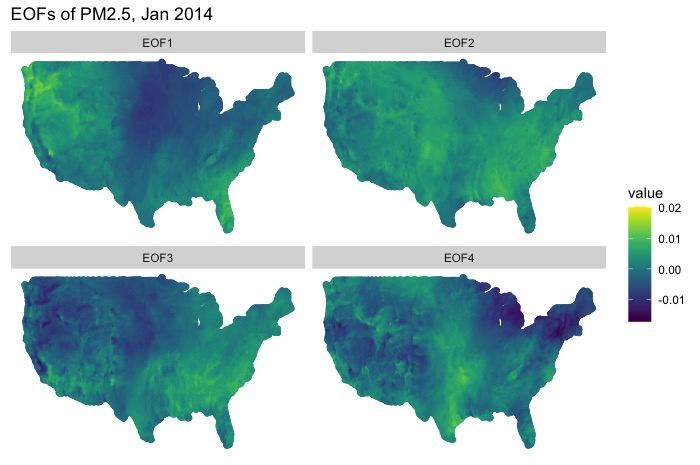}, 
\caption{\label{fig:eof} Leading EOFs of CMAQ output for January 2014. }
\end{figure}

In this paper, we derive the leading EOFs from multiple days of CMAQ data to construct a reduced set of explanatory variables that capture the dominant spatial patterns present in the numerical model output. Figure \ref{fig:eof} displays the spatial pattern of the first 4 EOFs, derived from CMAQ data for January 2024.

\subsection{Spatial Model}

In this subsection, we introduce the spatial model with a mean component based on EOFs of the CMAQ data. Let $y_t(\bs)$ represent the response observed from AQS monitor network on day $t$ at spatial location $\bs\in\mathcal{D}\subset\mathbb{R}^2$. The model is defined as 
\begin{equation}\label{e:y}
  y_t(\bs) = \mu_t(\bs) +  w(\bs) + \varepsilon_t(\bs)
\end{equation}
where $\mu_t(\bs)$ denotes the mean effect account for the macro-spatial structure and temporal trend, $w(\bs)$ is the spatial process modeling local small-scale spatial variability, and $\varepsilon_t \iid \normal(0,\tau^2)$ is the error term, independent of $w(\bs)$, with local nugget variance $\tau^2$. The spatial process $w(\bs)$ is typically modeled as a Gaussian process, $\bw(\cdot) \sim \GP(\bfzero,\bC)$, with mean $\bfzero$ and spatial covariance matrix $\bC$.

The mean component $\mu_t(\bs)$ is modeled as a spatial regression onto the first $M$ EOFs derived from time-replicated CMAQ data, along with $x_t(\bs)$, the CMAQ output for the grid cell that containing $\bs$. This part captures the large-scale spatial pattern of PM$_{2.5}$ concentrations over time.  Explicitly, the mean component is written as
\begin{equation}\label{e:mu}
\mu_t(\bs) = \sum_{m=1}^{M}e_m(\bs)\beta_{t,m} + x_t(\bs)\beta_{t,M+1}.
\end{equation}
Here, $e_m(\bs)$ is an element of the $m$-th EOF that corresponds to location $\bs$, and $\beta_{t,m}$ for $m \leq M$ is the universal coefficient of the $m$-th EOF.\\

The process $\bw(\cdot)$ models the spatial variability, which is captured using a non-stationary Mat\'ern covariance function following \citet{Stein2005} and \citet{Paciorek2006}. It allows the correlation structure of PM$_{2.5}$ concentrations to vary across space. The covariance between two locations $\bs_i$ and $\bs_j$ is expressed as

\begin{equation}\label{e:C}
C(\bs_i,\bs_j) = \frac{\sigma_i\sigma_j}{\Gamma(\nu_{ij} )2^{\nu_{ij} -1}}|\Sigma_i|^{\frac{1}{4}}|\Sigma_j|^{\frac{1}{4}}|\frac{\Sigma_i+\Sigma_j}{2}|^{-\frac{1}{2}}(2\sqrt{\nu_{ij}Q_{ij}})^{\nu_{ij}}  \mathcal{K}_{\nu_{ij}} (2\sqrt{\nu_{ij} Q_{ij}})
\end{equation}
where $\nu_{ij}= (\nu_i + \nu_j)/2$. The quadratic form $Q_{ij}$ is defined as

\begin{equation}\label{e:Q}
Q_{ij} = (\frac{\Sigma_i+\Sigma_j}{2})^{-1} (\bs_i-\bs_j)^T(\bs_i-\bs_j),
\end{equation}
where \(\sigma^2_i\) is the marginal variance at location \(\bs_i\), and \(\mathcal{K}\) denotes the modified Bessel function of the second kind. The parameters \(\nu_i\) and \(\rho_i\) control the local smoothness and spatial range of the process, respectively. The matrix \(\Sigma_i\) is defined as \(\Sigma_i = \Sigma_{A(i)}\), where \(A(i)\) denotes the CMAQ-defined region containing location \(\bs_i\). Each \(\Sigma_{A(i)}\) is a symmetric positive definite matrix that characterizes the local covariance structure of the Gaussian kernel centered at \(\bs_i\), and encodes the spatial parameters within region \(A(i)\).\\

The model extends Matérn covariance function to a non-stationary setting, by incorporating local $\Sigma_i$ matrices, which captures variations in spatial correlation strength. This flexibility allows spatial covariance structures to vary smoothly across locations, making it highly useful for modeling environmental and air pollution data, where the spatial correlation of pollutants like PM$_{2.5}$ often exhibits nonstationary behavior due to factors such as terrain, meteorology, and human activity.  However, the introduction of local covariance matrices can lead to identifiability issues, particularly when estimating spatially varying smoothness parameters \citep{Stein2005}. To mitigate the challenge, our study assume a constant smoothness parameter, $\nu_{ij} = \nu$. Additionally, to reduce model complexity and improves identifiability, we assume that the 2x2 kernel matrices have the form $\Sigma_i = \rho_i\mathbf{I}_2$, which implies that the process is locally isotropic with range parameter $\rho_i$. This implies that spatial correlation decays equally in all directions around each location, limiting the model's ability to capture anisotropic spatial dependence While this simplification prevents the model from capturing anisotropic spatial dependence, it remains effective for modeling spatially varying correlation range and tends to be more stable in practice, especially when data are sparse or directional structure is weak \citep{Paciorek2006}. \\

Although non-stationary models offer a more accurate representation of spatial variability, the increased flexibility comes at a computational cost and fitting the model to large datasets can become challenging. To address this, we implement a localized transfer learning approach to enhance computational efficiency. \\

\subsection{Localized transfer learning}

Transfer learning, originally developed in the field of machine learning \citep{Pan2010}, involves re-purposing a model trained on one task for another related task. In our case, we leverage the numerical CMAQ model output and fit models to estimate the local spatial parameters and then use these estimates as priors for the AQS model. \\

For the non-stationary covariance function of AQS in \eqref{e:C}, a set of spatial varying parameters, $\bftheta_i = [\rho_i,\sigma_i$], must be estimated. Ideally, these parameters would be directly estimated from the AQS monitoring data. However, due to the sparse coverage of AQS data, estimating these parameters directly is challenging. Instead, we estimate the parameters using a log-linear regression model based on the estimates from CMAQ output. \\

Let $\hat{\bftheta}_i=[\hat{\rho}_i,\hat{\sigma}^2_i]^T$ represent the estimates from the CMAQ output for the region where $\bs_i$ is located. To obtain $\hat{\bftheta}_i$ from the CMAQ output, we employ a moving window approach. We first remove the dominant large-scale spatial structure from the CMAQ output by projecting the data onto the leading EOFs and subtracting the resulting mean field. The residual field is then used to estimate local covariance parameters. The spatial domain was then divides into varying sizes (e.g., 1×1, 2×2, and 3×3 degrees), forming a moving window. Within each window, CMAQ outputs are treated as time independent replicates to estimate the local spatial parameters, \(\hat{\rho}_i\) and \(\hat{\sigma}^2_i\), for that region. This assumption simplifies the estimation by allowing us to pool temporal information while focusing on the spatial structure. The parameters are obtained using maximum likelihood estimation (MLE) under a locally stationary Matérn covariance model, applied to the CMAQ data within each window.\\

The moving window approach ensures that spatial heterogeneity across the domain is accounted for, as each window allows for the estimation of locally relevant parameters. Regions with higher spatial variability, such as urban areas, naturally capture shorter spatial ranges and higher variances, while smoother regions like rural areas exhibit longer spatial ranges and lower variances. This localized parameterization is critical for constructing the non-stationary covariance function in \eqref{e:Cc}, as it allows the model to reflect region-specific spatial correlation structures.  \\

Using $\hat{\bftheta}_i$, we then assign priors in \eqref{e:C} through the following log-linear model:
\begin{equation}\label{e:logreg}
\begin{aligned}
\log(\rho_i) &= a_1 + b_1 \log(\hat{\rho}_i) \\
\log(\sigma^2_i) &= a_2 + b_2 \log(\hat{\sigma}^2_i)
\end{aligned}
\end{equation}
The primary advantage of this model is the reduction in the number of parameters that need to be estimated in the Bayesian framework. Instead of estimating all local parameters, we estimate only the two sets of coefficients, $\ba$ and $\bb$. Notably, 
the model simplifies to a stationary model if $b_1=b_2=0$, which ignoring the CMAQ correlation. 
 In addition, the CMAQ model outputs are provided on a fixed grid, whereas the in-situ AQS observations are point-referenced. To address the change-of-support issue, we match each observation location to its nearest CMAQ grid cell.



\subsection{Bayesian Hierarchical Model}\label{subsec:bhm}
In this section, we develop the fully Bayesian framework for estimating the parameters of the spatial model described earlier. The goal of the Bayesian approach is to infer both the mean and spatial process effect while quantifying uncertainty in the model's estimates. The Bayesian hierarchical model is structured in three primary levels.\\

The first level specifies the data model, which defines the distribution of the observed data, conditional on the mean component and the spatial process. We assume the following for the response
\begin{equation}
  y_t(\bs) | \mu_t(\bs), w(\bs)  \indep  \normal(\mu_t(\bs) + w(\bs),\tau^2).
\end{equation}
The second level defines the underlying process, conditional on the model parameters. The mean component and the spatial process are
\begin{equation}
\begin{aligned}
\mu_t(\bs) & = [E(\bs),x_t(\bs)]^{T}\bfbeta_t,\\ 
\bw(\cdot)|\nu, \rho(\cdot),\sigma(\cdot) & \sim \GP(\bfzero,C(\cdot,\cdot)),
\end{aligned}
\end{equation}
where $E(\bs)$ is the matrix for first $M$ EOFs derived from time-replicated CMAQ data. The covariance function $\bC$ is specified as  
\begin{equation}\label{e:Cc}
C(\bs_i,\bs_j)  = \sigma_i\sigma_j \mathcal{M}_{ns}(\bs_i,\bs_j, \nu, \rho_i,\rho_j),
\end{equation}
where $\mathcal{M}_{ns}$ represents the non-stationary version of Mat\`ern covariance function in \eqref{e:C}. The local parameters are modeled as \eqref{e:logreg}, in which
$\hat{\rho}_i$ and $\hat{\sigma}^2_i$ are the regional parameter estimates derived from the CMAQ output using a moving windows approach.\\

The third level addressed the uncertainty in the parameters by specifying non-informative prior distributions. 
\begin{equation}
\begin{aligned}
\beta_{t,m} & \sim \normal(0, \omega^2),\\
\tau^2, \omega^2 & \sim \textit{InvGamma}(0.1,0.1),\\
a_1,a_2,b_1,b_2 & \sim \normal(0,10^2),\\
\nu & \sim \textit{Uniform}(0,3).
\end{aligned}
\end{equation}
Here, $\beta_{t,m}$ are the time-varying regression coefficients for the mean part, $\sigma^2$ represents the error variance, and $\nu$ controls the smoothness of the spatial process. The use of non-informative priors reflects our intent to let the data inform most of the model's inferences.

\section{Simulation study \label{sec:sim}}
\begin{figure}
\centering
\includegraphics[ width=.7\textwidth]{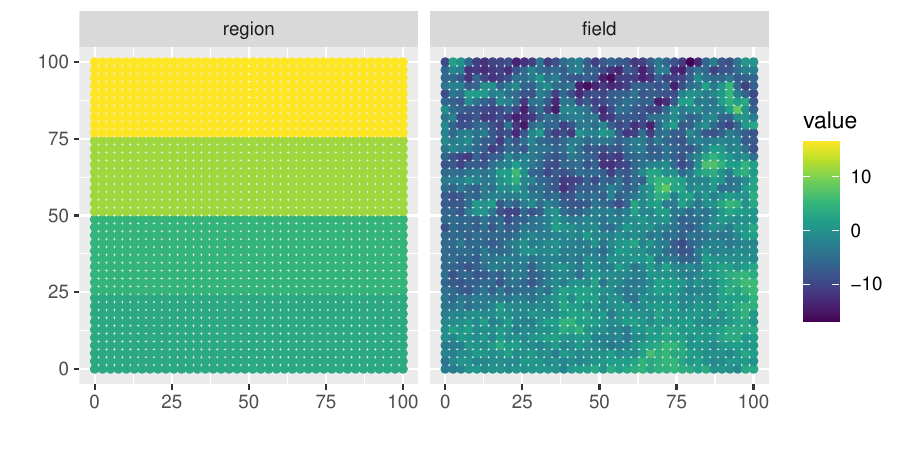}
\caption{\label{fig:sim_z} Regions with different parameter sets and simulated field}
\end{figure}

\begin{figure}
\centering
\includegraphics[ trim = 7mm 7mm 4.5mm 0mm, clip, width=.95\textwidth]{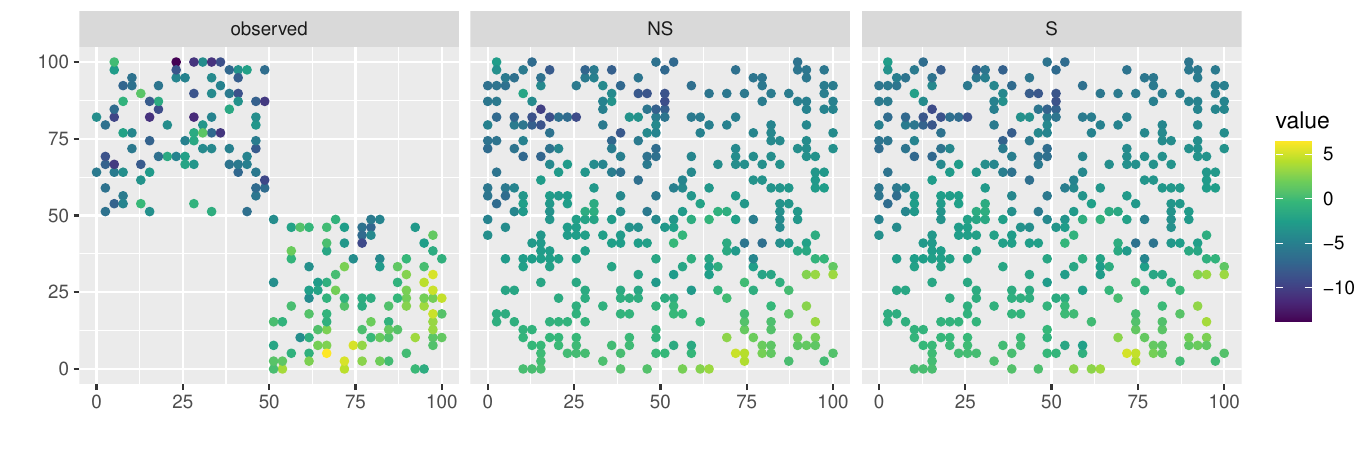}
\caption{\label{fig:sim_pd} Sampled observations and comparison of posterior prediction between the non-stationary model (NS) and the stationary model (S) for day $t=3$.}
\end{figure}

In this section, we use simulated data to compare the non-stationary spatial transfer learning model against a stationary model. The underlying field is simulated on a 40x40 grid over a four-days period in a two dimensional domain $\domain = [1,100]^2$. Its mean is defined by two explanatory variable, the longitude and latitude of the grid locations.  For $t = 1,\ldots,4$, the mean at $\bs=(s_1,s_2)$ is given by
\begin{equation}
\mu_t(\bs) = 0.05 s_1 + 0.1 s_2
\end{equation}
To simulate a non-stationary spatial process, the domain is divided into four horizontal sub-regions as illustrated in the top left panel of Figure \ref{fig:sim_z}. Each region has its own spatial range parameter $\rho$ and marginal variance $\sigma^2$, as specified in  Equation \eqref{e:Cc}. To evaluate model performance across different data densities, the domain is further divided into four corner sections, with two corners have no observation (all-missing) and two with partial observed data (partial-missing). Each scenario spans all four horizontal sub-regions. The observations are randomly sampled from two sub-regions on the upper left and lower right corners of the simulated field, therefore these sub-regions are partially missing. And the other two corner regions are all-missing regions, as shown in the left panel of Figure \ref{fig:sim_pd}. \\

To mimic the CMAQ model output in real data, a reference spatial field is simulated with base parameters $\bftheta_0 =[\bfrho_0 \quad  \bfsigma^2_0]$. And $\bfrho_0 = [15,10,5,2.5]^T$ are the spatial range parameters for the 4 regions, $\bfsigma^2_0 = [2, 3, 7, 10]^T$ are their marginal variances. The universal smoothness parameter $\nu=1.5$ is applied across all regions. The relationship between the spatial parameters of the true underlying field, $\bftheta$, and the parameters of the reference field, $\bftheta_0$, is assumed to follow
\[log(\bftheta)  = 0.5 + log(\bftheta_0).\]

\begin{figure}
\centering
\includegraphics[ width=.7\textwidth]{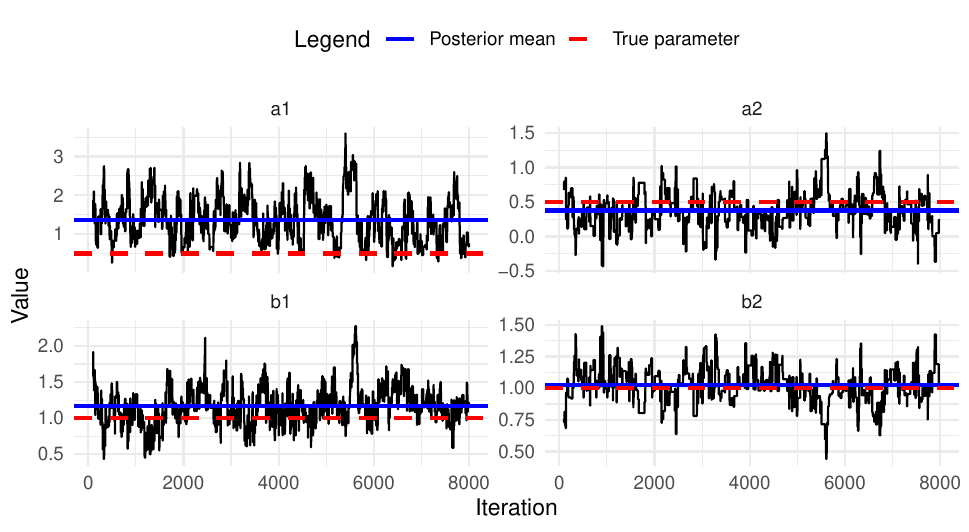}
\caption{\label{fig:trace} Trace plots of the posterior estimates for parameter $a1,b1,a2,b2$.}
\end{figure}

We generated 50 datasets by adding independent normal measurement error with variance 1. For each dataset, we divide it into training and testing data, then ran 10,000 MCMC iterations on training data with the first 15\% discarded as burn-in (determined by trace plots from a pilot study). The prior distributions of the parameters are described in Section \ref{subsec:bhm}. The convergence of chains are checked and the trace plot in Figure \ref{fig:trace} indicate that the posterior estimates of $b_1,a_2,b_2$ are closely aligned with the true parameters used in the simulation. While the estimate of $a_1$ exhibits some bias, it still captures the correct direction and magnitude of adjustment relative to the CMAQ-derived inputs.This suggests that the Bayesian model effectively captures the relationship between the observed field and the reference field.\\ 

\begin{figure}
\centering
\includegraphics[ width=.6\textwidth]{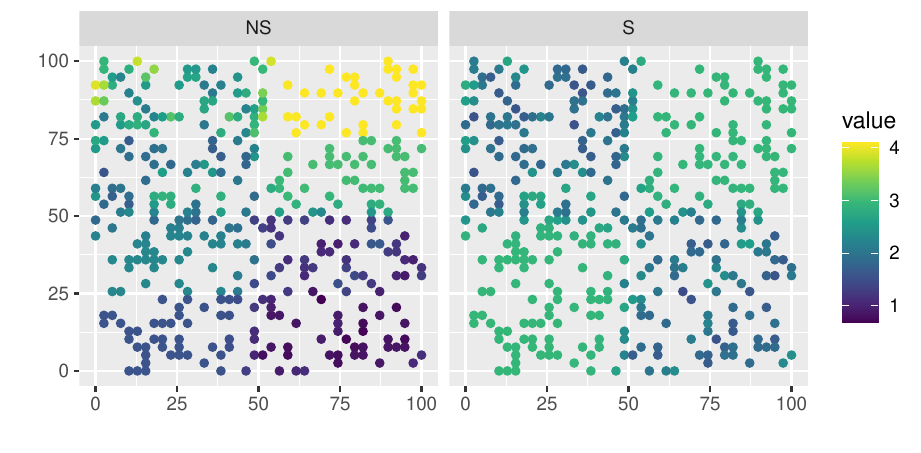}
\caption{\label{fig:sim_pdse} Comparison of the prediction standard errors for the non-stationary model (NS) and stationary model (S) for day $t=3$.}
\end{figure}

Posterior predictions and their standard errors were recorded for both non-stationary (NS) and stationary (S) spatial models using testing data. Figure \ref{fig:sim_pd} illustrates the comparison of predictions on testing data alongside training observations, while Figure \ref{fig:sim_pdse} presents the prediction standard errors for both models. Generally, the non-stationary model provides more accurate uncertainty quantification. Prediction error at each location is computed as the squared difference between the predicted posterior mean and the true value, i.e., \(\left(\hat{y}_t(\bs) - y_t(\bs)\right)^2 \). Notably, the non-stationary model exhibits lower prediction errors in the bottom two regions where predictions are more accurate and higher standard errors in the top region where predictions deviate significantly from the true field.\\ 

Table \ref{tab:sim} summarizes the results of the simulation study. We compared the non-stationary model with the stationary model under the transfer learning framework using several metrics: continuously rank probability score (CRPS), negative log-likelihood as log-loss, and root mean squared prediction error (RMSE).  Across all metrics, the non-stationary model consistently outperforms the stationary model, primarily due to its superior uncertainty quantification. This is evidenced by lower standard errors for more accurate predictions and higher standard errors for less accurate ones. However, in terms of prediction accuracy, as measured by RMSE, the non-stationary model shows only a marginal advantage. Predictions in areas with partial data missing generally perform better than those in completely missing areas.\\

\begin{table}[htbp]
\centering
\small
\begin{tabular}{l|l|rrr}
Metric & Model & Partial-missing & All-missing & Overall\\ 
\hline
\multirow{2}{*}{log-loss} 
&  NS &  \textbf{4396} & \textbf{5509} & \textbf{9905}  \\     
& S  & 64503 &  64455 & 129058 \\ 
\hline
\multirow{2}{*}{CRPS} 
  & NS & \textbf{1028} & \textbf{1436}  & \textbf{2464}   \\ 
  & S & 1092  & 1487 & 2579 \\ 
\hline
\multirow{2}{*}{RMSE} 
 &  NS & \textbf{2.327} & \textbf{3.226} & \textbf{2.815} \\  
 & S  &  2.370  &  3.232 & 2.838\\
\end{tabular}
\caption{Model comparison with different metrics for simulated data, log-loss: negative log-likelihood. CRPS: continuous rank probability score. RMSE: root mean squared prediction error.} 
\label{tab:sim}
\end{table}

\section{Application \label{sec:app}}

In this section, we apply the proposed non-stationary spatial model to the AQS-observed PM$_{2.5}$ data in the lower 48 states of the United States, as introduced in Section \ref{sec:data}. Let $y_t(\cdot), t=1, \ldots, 3$, represent the first three day PM$_{2.5}$ measurements from AQS observations in January 2014. The model is formulated as
\begin{equation}
  y_t(\bs) = \sum_{m=0}^{7}e_m(\bs)^T\beta_{t,m} + x_t(\bs)^T\beta_{t,8} +  w(\bs) + \varepsilon_{t,s}
\end{equation}
where the mean component is modeled as a spatial regression on the first 7 EOFs, which are decomposed from CMAQ data for January 2014. The term $x_t$ is the CMAQ output for the same date $t$ as the response. The 7 leading EOFs explain about $70\%$ of the variance in the CMAQ data, effectively capturing large-scale patterns of PM$_{2.5}$ over the domain. We assume that large-scale spatial variability remains relatively stable over time, while the time-varying coefficient $\beta_{t,m}$ account for temporal trend in the mean. The spatial process $w(\cdot)$ accounts for the remaining small-scale spatial correlation, and it is modeled using a non-stationary Mat\'ern covariance function as defined in Equation \ref{e:Cc}. The covariance matrix $\Sigma_i$ in Equation \ref{e:C} is constructed as a diagonal matrix, dependent solely on the spatial range of each local region.\\  

As an input to the non-stationary covariance function in Equation \ref{e:C}, we estimate a set of spatial varying parameters, $\bftheta_i = [\rho_i,\sigma_i$] using the pre-estimated spatial parameters $\hat{\bftheta}_i$ based on CMAQ output. To obtain \(\hat{\bftheta}_i\), we first remove the dominant large-scale spatial structure from the CMAQ output by projecting the data onto the leading EOFs and subtracting the resulting mean field. The residual field is then used to estimate local covariance parameters. Specifically, the domain is divided into adjacent square regions of varying sizes (e.g., 1×1, 2×2, and 3×3 degrees), forming a moving window. Within each square region, spatial parameters are estimated using a Gaussian process model with a stationary Matérn covariance function applied to time-replicated residuals from the EOF-adjusted CMAQ data, which is assumed independence across time points. The local parameter estimation within each region is carried out with R package \texttt{geoR} \citep{geoR}.\\

\begin{figure}[tbp]
\centering

\includegraphics[trim = 0cm 0cm 3cm 0cm, clip, width=.3\textwidth]{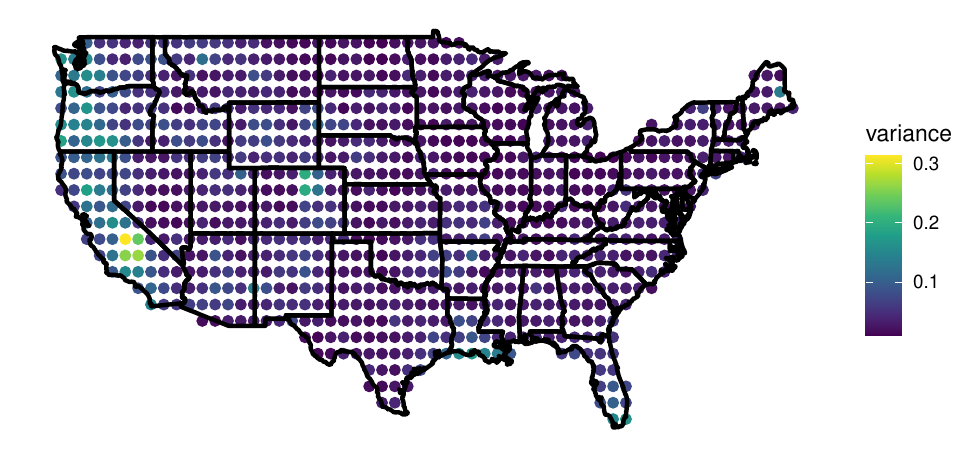}
\includegraphics[trim = 0cm 0cm 3cm 0cm, clip, width=.3\textwidth]{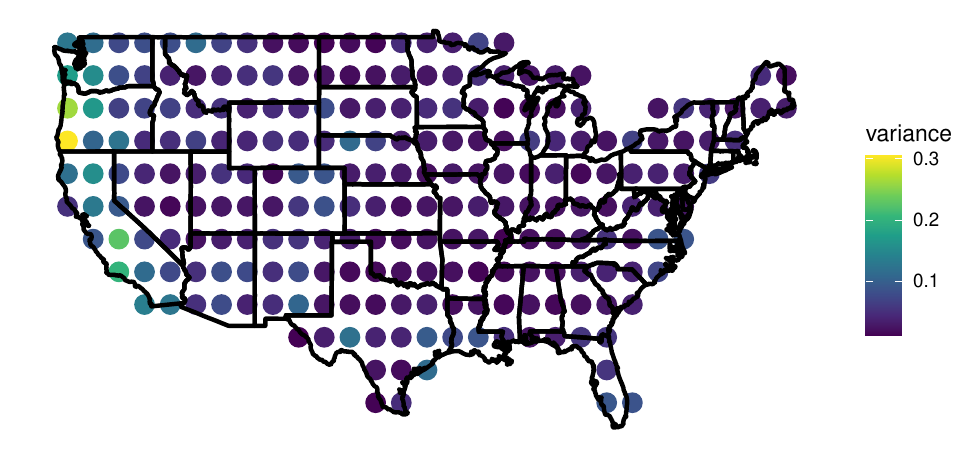}
\includegraphics[ width=.35\textwidth]{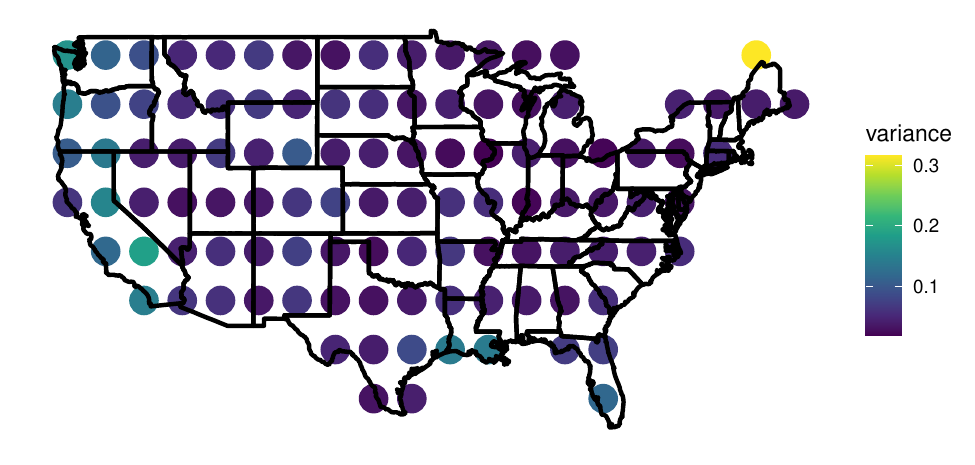}

\includegraphics[trim = 0cm 0cm 3cm 0cm, clip, width=.3\textwidth]{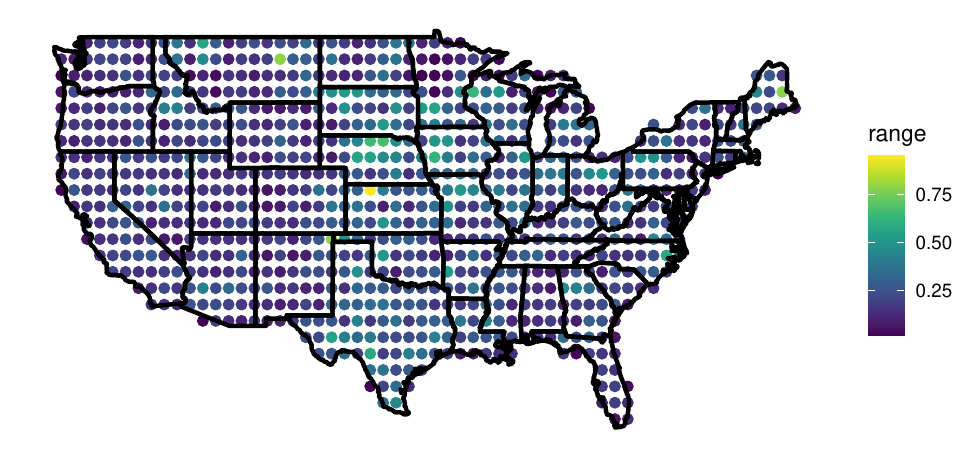}
\includegraphics[trim = 0cm 0cm 3cm 0cm, clip,  width=.3\textwidth]{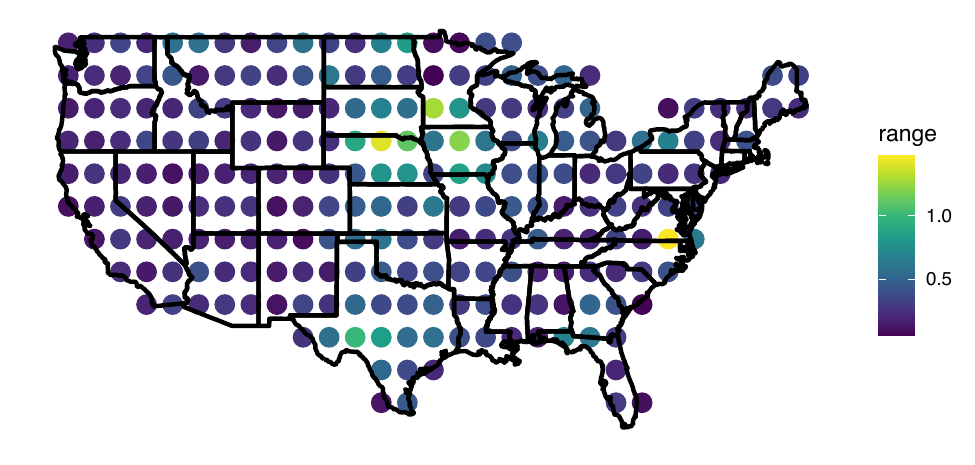}
\includegraphics[width=.35\textwidth]{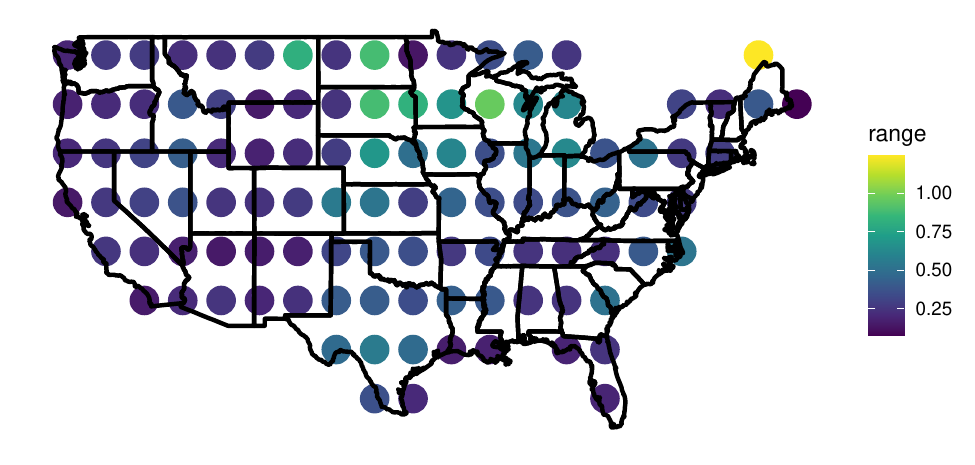}

\caption{\label{fig:param}  Top: moving window estimates of the marginal variance  parameter based on adjusted CMAQ data with 1x1,2x2 and 3x3 degree grid from left to right. Bottom: moving window estimates of the spatial range parameter based on adjusted CMAQ data with 1x1,2x2 and 3x3 degree grid from left to right.}
\end{figure}

To assess the sensitivity of local parameter estimates to the choice of window size, we repeated the estimation using different grid resolutions. Finer grids (e.g., 1×1) capture more localized spatial variability, while coarser grids (e.g., 3×3) smooth out small-scale fluctuations and emphasize broader regional trends. Figure \ref{fig:param} displays the spatial range and marginal variance parameters of using the moving windows approach. The top panel of Figure \ref{fig:param} shows that the east coast has generally higher marginal variances, while the bottom panel reveals that the middle region exhibits relatively large spatial ranges. These broad spatial patterns remain qualitatively consistent across grid sizes, though the finer windows reveal more pronounced local heterogeneity.\\

The estimation of $\bftheta_i$ is performed using a simple linear regression on logarithmic scale, as shown in Equation \ref{e:logreg}. 
The parameters $a_1,a_2, b_1, b_2$ are estimated within a Bayesian framework and the MCMC chains exhibit good mixing and convergence. The intercept  $a_1$ and $a_2$ converge near 3 (with a standard deviation (sd) of 0.3 for $a_1$ and 0.4 for $a_2$), suggesting a systematic offset between the CMAQ reference field and the observed AQS data. This implies that, on average, the parameters derived from CMAQ underestimate those from observed $PM_{2.5}$ concentrations, necessitating an upward adjustment. Similarly, the slope parameters $b_1$ and $b_2$ converge close to 0.8 (sd=0.3) and 1.2 (sd=0.1) respectively, indicating a differential scaling effect between the reference field and the observed data. The slopes $b_1$ and $b_2$ are positive with probability close to one which establishes that there is an association between the CMAQ-derived parameters and those from the AQS. Since $a_1$ and $b_1$ correspond to modeling the range parameter, the fact that $b_1 <1$ suggests that CMAQ tends to overestimate the spatial range, effectively leading to oversmoothing in space. Meanwhile, $a2$ and $b_2$ which govern marginal variance, indicate that $b_2 \approx 1.18$ reflects a slight amplification of observed variations. These results underscore the importance of transfer learning in correcting biases and scaling mismatches between CMAQ outputs and ground observations, allowing the model to better adapt to spatial heterogeneity in air pollution data. All code and reproducible materials will be posted online after the manuscript is published.\\



 The dataset is divided spatially into 90\% training locations and 10\% held-out test locations. Both models are fitted to each training set and used to make out-of-sample spatial predictions at the corresponding test locations. We perform 10-fold cross-validation over spatial locations to ensure that all data points are used for both training and testing. Since the split is spatial rather than temporal, our evaluation focuses on spatial interpolation rather than temporal forecasting. The predictive performance of the non-stationary (NS) model is compared to that of the stationary (S) model as well as a baseline model with no spatial terms (no spatial) for the application data, as summarized in Table \ref{tab:app}. The comparison is based on four key metrics: negative log-likelihood as log-loss, continuously ranked probability score (CRPS), and root mean squared error (RMSE) and 95\% coverage rate (95CR). And all the metrics are summed over the 10 testing datasets. For the NS model, we further consider the three different grid settings (1×1, 2×2, and 3×3 degrees) to evaluate the sensitivity of predictive performance to the choice of spatial resolution. These parallel comparisons allow us to assess whether finer grids improve local prediction accuracy or whether coarser grids provide more stable estimates.\\
 
As shown in Table \ref{tab:app}, the no-spatial model performs substantially worse than both the stationary and nonstationary spatial models across all scoring metrics (beside a higher coverage rate likely from its larger prediction variance). This indicates that low-rank CMAQ information alone is inadequate for accurate prediction and that the dominant improvements arise from explicit spatial modeling and the proposed nonstationary transfer-learning covariance. We also examined sensitivity to the number of EOF modes (3, 5, 7, and 8) and found that performance stabilizes around 7 modes; results beyond this point showed negligible improvement and are therefore not reported in detail.



\begin{table}[htbp]
\centering
\small
\begin{tabular}{c|c|c|ccc}
 & no spatial & S & \multicolumn{3}{c}{NS} \\
 & &   & 1x1 & 2x2 & 3x3 \\
 \hline 
log-loss   & 849.5 & 453.1 & 396.1 & \textbf{392.3} &  414.0 \\ 
\hline
CRPS     & 298.3 & 175.8 & 173.0 & \textbf{171.5} &  173.9 \\ 
\hline
RMSE   & 0.768 & 0.480 & \textbf{0.479} & \textbf{0.479} &  0.481 \\
\hline
95CR   & \textbf{95.9\%}  & 92\% & 87\% & 89\%  & 92\% \\
\end{tabular}

\caption{Model comparison for application data: model with no spatial term vs. with stationary spatial covariance (S)  vs. non-stationary spatial model (NS)  with 1x1, 2x2 and 3x3 degree grid.} 
\label{tab:app}
\end{table}

\begin{figure}[ht]
\centering
\includegraphics[ width=.5\textwidth]{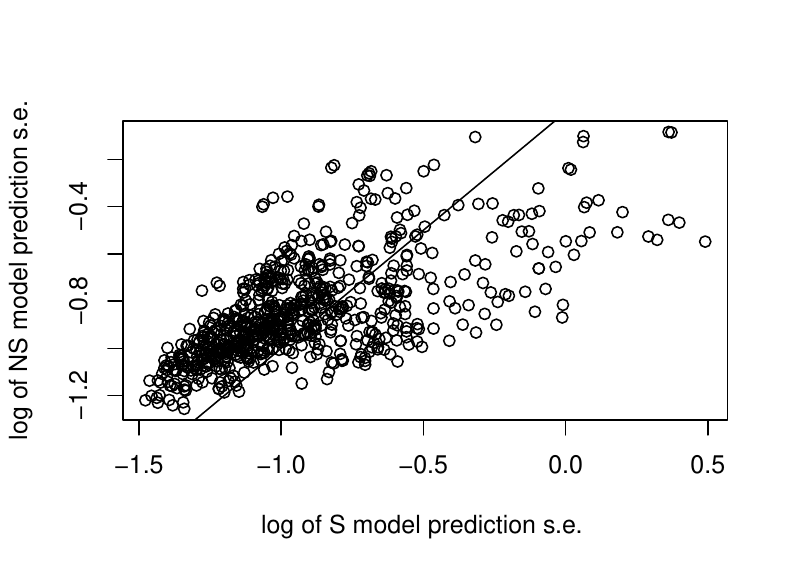}
\caption{\label{fig:pdse}  Comparison of the posterior prediction standard error for testing data (with 2x2 degree grids).}
\end{figure}

The results indicate that the non-stationary model outperforms the stationary model in terms of log-loss, CRPS, and RMSE. The improvement is most pronounced for log-loss and CRPS, while the reduction in RMSE is relatively modest. In contrast, the stationary model achieves a higher coverage rate, due to its larger prediction standard errors (Figure \ref{fig:pdse}). Overall, despite the marginal difference in RMSE, the non-stationary model provides a more faithful representation of the underlying spatial variability in the PM$_{2.5}$ data. This finding aligns with \citet{Paciorek2006} and other studies, which suggest that while non-stationary models often show limited gains in predictive accuracy or coverage over stationary counterparts, they tend to offer improved uncertainty quantification.\\

When examining the effect of different grid sizes, the finer-resolution NS models (1×1 and 2×2 degrees) yield the strongest overall performance, with the 2×2 grid producing the lowest log-loss and CRPS. By contrast, the coarser 3×3 grid results in less precise predictions and inflated standard errors, producing coverage rates closer to those of the stationary model. These results suggest that finer grids enable the non-stationary model to better capture local spatial variability, whereas coarser grids tend to over-smooth and reduce predictive performance.

\section{Discussion \label{sec:conclusions}}
This study introduces a novel approach to modeling PM$_{2.5}$ concentrations by integrating numerical model outputs with observed monitoring data through a non-stationary spatial transfer learning framework. The model leverages both Empirical Orthogonal Functions (EOFs) derived from the CMAQ data and a non-stationary Matérn covariance function to account for both large and small scale spatial variability. The use of EOFs to capture dominant spatial patterns in PM$_{2.5}$, while the non-stationary Matérn covariance function provides flexibility needed to model regional spatial variations.\\

By incorporating the numerical model as a pre-trained framework for spatial parameter learning, we effectively bridge the gap between sparse observational data and the more comprehensive numerical output, significantly reducing the number of parameters that need to be estimated. A key advantage of the non-stationary model is its ability to adapt to varying spatial correlation structures across different regions. This flexibility is particularly valuable in environmental applications, where reliable exposure assessments are crucial despite the limitations of sparse monitoring data. \\

While the improvement in point prediction accuracy over the stationary model is modest, this is consistent with findings in \citet{Paciorek2006}, which noted limited quantitative gains in prediction from nonstationary models. The main benefit of the nonstationary approach lies in its improved uncertainty quantification. Both the simulation study and real-world applications highlight this strength, as demonstrated by the reduced prediction variance and superior performance on evaluation metrics such as log-loss and CRPS. Our sensitivity analyses, including comparisons across different moving window sizes, further confirm the robustness of these findings.\\

A limitation of the current framework is that uncertainty in CMAQ-driven parameters is not explicitly propagated. Instead, uncertainty is incorporated through the transfer learning calibration stage, where the regression parameters linking CMAQ to AQS are estimated within the hierarchical model and their posterior uncertainty is naturally carried forward into prediction. While this partially addresses uncertainty propagation, more comprehensive Bayesian data assimilation frameworks could further improve robustness by jointly modeling these parameters.\\

Our model assumed the dominant spatial pattern extracted in EOFs are invariant over time but used their coefficients to account for temporal trend. Expanding the framework to handle spatio-temporal trend simultaneously in the non-stationary covariance model would be a valuable extension, allowing for better predictions over time while accounting for both spatial and temporal dependencies. The computational complexity of the non-stationary model is slightly higher than that of the stationary one as estimating the local parameters is necessary. Despite the reduction in parameter estimation complexity achieved through localized transfer learning, there remains a risk of overfitting, particularly in regions with very sparse data coverage. Future research could explore more efficient computational techniques, such as variational inference, to reduce the computational burden while preserving the flexibility of the non-stationary model. Additionally, incorporating more sophisticated regularization methods could help mitigate the risk of overfitting and improve the model's performance in extremely sparse data scenarios.

\footnotesize
\appendix
\section*{Acknowledgments}

The authors would like to thank Daniel Zilber for helpful comments and discussions and we also appreciate the effort of the Editor, Associate Editor and Referees to improve quality of the manuscript.


\bibliographystyle{apalike}
\bibliography{mendeley,additionalrefs}

\end{document}